\documentclass[aps,prx,longbibliography,twocolumn,floatfix,superscriptaddress]{revtex4-1}
\usepackage{dcolumn,amsmath}
\usepackage{graphicx}
\usepackage{braket}
\usepackage{bm}
\usepackage{amssymb}
\usepackage{hyperref}
\usepackage{multirow}
\usepackage{footnote}
\usepackage{siunitx}
\usepackage{subcaption}
\usepackage{ragged2e}
\usepackage{xcolor}
\usepackage{pdfcomment}
\usepackage[justification=justified,labelfont=bf,singlelinecheck=off]{caption}
\usepackage{threeparttable,booktabs} 
\usepackage{subcaption,siunitx,booktabs}
\usepackage{caption,afterpage,tabularx}
\usepackage{diagbox}
\usepackage{csquotes}

\newcommand{\RUG}{
Van Swinderen Institute for Particle Physics and Gravity,
Faculty of Science and Engineering, \\University of Groningen, Nijenborgh 4, 9747 AG Groningen, The Netherlands\\
}
\newcommand{\MBU}{
Department of Chemistry, Faculty of Natural Sciences, 
\\Matej Bel University, Tajovsk{\'e}ho 40, 97401 Bansk{\'a} Bystrica, Slovakia\\
}

\newcommand{\TAU}{School of Chemistry, 
Tel Aviv University, 6997801 Tel Aviv, Israel
}

\newcommand{\UNSW}{School of Physics, 
University of New South Wales, Sydney, New South Wales 2052, Australia
}

\newcommand{\NIST}{Joint Quantum Institute, National Institute of Standards and Technology and University of Maryland,
Gaithersburg, Maryland 20899, USA}

\newcommand{\TRIUMF}{TRIUMF, 4004 Wesbrook Mall, Vancouver, British Columbia V6T 2A3, Canada}

\begin{document}

\title{Nuclear spin-dependent parity-violating effects in light polyatomic molecules}

\author{Yongliang Hao}
\affiliation{\RUG}
\author{Petr Navr\'{a}til}
\affiliation{\TRIUMF}
\author{Eric B. Norrgard}
\affiliation{\NIST}
\author{Miroslav Ilia\v{s}}
\affiliation{\MBU}
\author{Ephraim~Eliav}
\affiliation{\TAU}
\author{Rob~G. E. Timmermans}
\affiliation{\RUG}
\author{Victor V. Flambaum}
\affiliation{\UNSW}
\author{Anastasia Borschevsky}
\email{a.borschevsky@rug.nl}
\affiliation{\RUG}

\date{\today}

\begin{abstract}
 
Measurements of nuclear spin-dependent parity-violating (NSD-PV) effects  provide an excellent opportunity to test nuclear models and to search for physics beyond the Standard Model. Molecules possess closely-spaced states with opposite parity which may be easily tuned to degeneracy to greatly enhance the observed  parity-violating effects. A high-sensitivity measurement of NSD-PV effects using light triatomic molecules is in preparation [E. B. Norrgard, \textit{et al.}, Commun. Phys. \textbf{2}, 77 (2019)]. Importantly, by comparing these measurements in light nuclei with prior and ongoing measurements in heavier systems, the contribution to NSD-PV from $Z^0$-boson exchange between the electrons and the nuclei may be separated from the contribution of the nuclear anapole moment.  Furthermore, light triatomic molecules offer the possibility to search for new particles, such as the postulated $Z^{\prime}$ boson. In this work, we detail a sensitive measurement scheme and present high-accuracy molecular and nuclear calculations needed for interpretation of NSD-PV experiments on triatomic molecules composed of light elements Be, Mg, N, and C. The \textit{ab initio} nuclear structure calculations, performed within the No-Core Shell Model (NCSM) provide a reliable prediction of the magnitude of different contributions to the NSD-PV effects in the four nuclei. These results differ significantly from the predictions of the standard single-particle model and highlight the importance of including many-body effects in such calculations. In order to extract the NSD-PV contributions from measurements, a parity-violating interaction parameter $W_{\text{PV}}$, which depends on molecular structure, needs to be known with high accuracy. We have calculated these parameters for the triatomic molecules of interest using the relativistic coupled-cluster approach. In order to facilitate the interpretation of future experiments we provide uncertainties on the calculated parameters. A scheme for measurement using laser-cooled polyatomic molecules in a molecular fountain is presented, along with an estimate of the expected sensitivity of such an experiment. This experimental scheme, combined with the presented state-of-the-art calculations, opens exciting prospects for a measurement of the anapole moment and the PV effects due to the electron-nucleon interactions with unprecedented accuracy and for a new path towards detection of signatures of physics beyond the Standard Model.
\end{abstract}

\maketitle

\section{Introduction}\label{Sec1}

Measurements and calculations of parity-violating effects in atoms and molecules are important both for the verification of the Standard Model (SM) and for the investigation of phenomena that cannot be explained within this model, such as the nature of dark matter and the matter-antimatter asymmetry. One of the candidates for the dark-matter particles is a low-mass $Z^{\prime}$ boson \cite{BouFay05,DavLeeMar12,DavLeeMar12-2}. The best limits on the parity-violating interaction of this $Z^{\prime}$ boson with electrons, protons, and neutrons were obtained from the data on atomic parity violation \cite{dzuba2017probing}; in particular, information on its interaction with nucleons was extracted from the measurements of the nuclear anapole moment of the $^{133}$Cs nucleus in Ref. \cite{WooBenCho97}. The possibility to study the nuclear anapole moments in additional systems, and thus to set further constraints on this interaction, provides a major motivation for the current work.

The notion of the anapole moment was introduced by Zel'dovich in 1958 \cite{Zel58}. The nuclear anapole moment was originally considered in Ref. \cite{FlaKhr80-1} and calculated in Ref. \cite{FlaKhrSus84} for a number of heavy atoms. This work also proposed possible schemes to observe nuclear anapole-moment effects in atomic and molecular experiments. Studies of the nuclear anapole-moment effects can provide information about parity-violating nuclear forces \cite{FlaKhr80-1,FlaKhrSus84} and may be considered as a test of nuclear theory and low-energy quantum chromodynamics. The nuclear anapole-moment rapidly increases with the nucleon number $A$ (as $A^{2/3}$) and dominates the nuclear-spin-dependent parity-violating (NSD-PV) effects in heavy atoms \cite{FlaKhr80-1,FlaKhrSus84}.  
 
In light atoms, additional parity-violating mechanisms become important or even dominate the nuclear anapole-moment effect. One such effect is the exchange of the Standard Model $Z^0$ boson (or potential yet-undiscovered $Z^{\prime}$ bosons) between an electron and individual nucleons \cite{NovSusFla77-1}, which remains poorly characterized despite the great strides recently in electron-scattering experiments \cite{WanPanSub14, Sou16}. Therefore, NSD-PV measurements in light nuclei are sensitive tests of the Standard Model and may be used to search for new particles such as $Z^{\prime}$ bosons and particles contributing to electroweak radiative corrections.

Compared with atoms, the NSD-PV effects are strongly enhanced in molecules
due to the close-lying states of opposite parity \cite{Lab78,SusFla78,FlaKhr85_2}. The Stark interference technique, which uses external fields to bring the rotational or hyperfine levels with opposite parity into near-degeneracy, has been widely employed in the search for these effects (see e.g. Refs. \cite{Nguyen1997,Tsigutkin2009,demille2008using,AltAmmCah18-1,AltAmmCah18-2}). A recent proposal identified linear triatomic molecules as promising systems to measure NSD-PV effects \cite{norrgard2019nuclear}. A general feature of such molecules is that they have closely-spaced $\ell$-doublets with opposite parity, allowing parity-violation-sensitive pairs of levels to be brought to degeneracy in magnetic fields typically two orders of magnitude smaller than needed for the diatomic molecules. Moreover, it is in principle possible to measure NSD-PV effects in all three nuclei of these molecules, which would allow the various underlying parity-violating effects to be disentangled.

Light triatomic molecules are especially attractive candidates for precision measurements of the NSD-PV effects. A proper interpretation of an NSD-PV measurement relies on accurate molecular and nuclear structure parameters. High-accuracy theoretical determination of the molecular properties becomes more computationally tractable for lighter systems, and, even more importantly, nuclear calculations are significantly more accurate and more reliable than in heavy elements. Here, we perform rigorous, high-accuracy calculations of the molecular and nuclear parameters required to interpret NSD-PV measurements in molecules composed of the light elements Be, C, N, and Mg.   We find that the parameters characterizing the molecule-specific sensitivity are in line with those of isoelectronic diatomic molecules \cite{borschevsky2013relativistic, hao2018nuclear}, as well as prior semi-empirical estimates \cite{Kozlov1985,norrgard2019nuclear}. However, our \textit{ab-initio} nuclear calculations find the nuclear anapole-moment interactions to be much stronger (typically 2 to 4 times larger) than predicted by a standard single-particle model \cite{FlaKhr80-1, FlaKhrSus84, FlaKhr85-1, FlaKhr85-2}, while the NSD-PV effects attributed to $Z^0$-boson exchange are typically reduced. This highlights the necessity of including many-body effects for correctly interpreting NSD-PV measurements, even in light nuclear systems. Moreover, the Be and Mg cyanide and isocyanide molecules considered here have favorable laser-cooling and trapping properties, which are essential to enabling high-sensitivity measurements through long interaction times.  We conclude by considering the experimental sensitivity to NSD-PV effects of laser-cooled molecules in free flight.  Using realistic parameters, the sensitivity of this method can exceed that of molecules in an optical trap \cite{norrgard2019nuclear}.

\section{Theory}

The NSD-PV interaction with the atomic or molecular electrons can be defined by the effective Hamiltonian \cite{FlaKhrSus84,Fla85}
\begin{eqnarray}
H_{\text{NSD-PV}}^{\text{eff}}=\frac{\kappa G_F}{\sqrt{2}} \Bigl(\frac{\bm{\alpha} \cdot \bm{I}}{I} \Bigr)\rho(\mathbf{r}),
\label{HNSD}
\end{eqnarray}
where $G_F$ is the Fermi weak interaction coupling constant, the Dirac matrices $\bm{\alpha}$ are defined in the usual way, $\bm{I}$ is the nuclear spin, and $\rho(\bm{r})$ is the nuclear density distribution function normalized to 1.

In a given nucleus, various underlying electroweak interactions  contribute to the total NSD-PV effect: \mbox{$\kappa = \kappa_\text{A} + \kappa_{\text{ax}} + \kappa_{\text{hfs}}$}.  In this section, we proceed by considering each of these three terms in turn, then explore how to evaluate Eq. (\ref{HNSD}) in a molecular system.

The effective coupling constant $\kappa_\text{A}$ describes the strength of the nuclear anapole-moment interaction. In a simple valence nucleon model, $\kappa_\text{A}$ takes the form \cite{FlaKhrSus84,Fla85}
\begin{equation}\label{eq:ka}
\begin{split}
  \kappa_{\text{A}} &= \frac{9}{10} \frac{\alpha \mu_i}{m_{p} r_0}g_i A^{2/3} \frac{K}{I+1}\\                   &\simeq 1.15\times 10^{-3} g_i \mu_i A^{2/3} \frac{K}{I+1}, 
\end{split}
\end{equation}
where $\alpha \simeq 1/137$ is the fine-structure constant, $m_{p}$ is the proton mass, $r_0 \simeq 1.2$\,fm is the scale of the nuclear radius, $\mu_i$ ($\mu_p\simeq2.793$ \cite{mohr2000codata,stone2005table} for proton, $\mu_n\simeq-1.913$ \cite{mohr2000codata,stone2005table} for neutron) is the nucleon magnetic moment in nuclear magnetons, $A$ is the mass number, and $K = (I+1/2)(-1)^{I-\ell_i+1/2}$, with $l_{i}$ being the orbital angular momentum (quantum number) of the external unpaired nucleon. The anapole contribution also depends on the poorly-known dimensionless constants $g_i$ ($i = p,n$), which characterize the nucleon-nucleus weak potential. In Refs. \cite{FlaKhrSus84,FlaMur97} these constants were expressed in terms of a meson-exchange model, and in Ref. \cite{FadFla19} the results based on different calculations of meson-nucleon interactions are presented. Using the most recent experimental data \cite{BlythFryFom18}, the authors of Ref. \cite{FadFla19} obtained $g_p=3.4 \pm 0.8$ and $g_n=0.9 \pm 0.6$. In the following, we will use central points $g_p=3.4$ and $g_n=0.9$ for the single-particle model estimates of the magnitude of the anapole moment. We note that this updated estimate of $g_n$ has opposite sign compared to the one used in earlier molecule NSD-PV considerations \cite{demille2008using,norrgard2019nuclear}. One of the aims of the measurements of  NSD-PV effects is to extract reliable values of these constants.

The nuclear anapole moment of $^{133}$Cs was  confirmed at a 7$\sigma$ significance level by Wood \textit{et al.}, with the value of $\kappa_\text{A} \simeq 0.392 \pm 0.056$ \cite{WooBenCho97}. A more accurate theoretical treatment performed after the experiment obtained a similar value \cite{FlaMur97}. Further NSD-PV measurements in Cs with improved precision have been proposed \cite{antypas2013measurement,ChoEli16}, and additional experiments have been designed to study the nuclear anapole-moment effect in other atoms with unpaired nucleons, such as $^{137}$Ba (using the BaF molecule) \cite{demille2008using}, $^{163}$Dy \cite{LeeGouAnt14}, $^{171}$Yb \cite{AntFabBou17}, and $^{212}$Fr \cite{AubBehCol13}.

The second contribution, $\kappa_\text{ax}$, is associated with the $Z^0$-exchange interaction between the electron vector and the nucleon axial-vector currents ($\bm{V}_e\bm{A}_N$) \cite{NovSusFla77-1}. The magnitude of $\kappa_\text{ax}$ within the nuclear shell model is \cite{FlaKhr80-1}
\begin{equation}\label{eq:kax}
\kappa_{\text{ax}} = C_{2} \frac{1/2-K}{I+1}, 
\end{equation}
where $C_{2}$ represents the $\bm{V}_e\bm{A}_N$ coupling, which takes the value $C_{2} \equiv -C_{\text{2p}}$ for a valence proton and $C_{2} \equiv -C_{\text{2n}}$ for a valence neutron \cite{ginges2004violations}. Here, $C_\text{2p}$ and $C_\text{2n}$ are given by
\begin{equation}
C_{\text{2p}} = -C_{\text{2n}} = g_A(1-4\sin^2\theta_W)/2 \simeq 0.05,
\end{equation}
with $g_A \simeq 1.26$ \cite{ginges2004violations} and sin$^2\theta_W \simeq 0.23$ \cite{Pat16}.

The PVDIS experiment  \cite{WanPanSub14} combined with the Cs PV measurement \cite{WooBenCho97} provides the best determination to date of the linear combination $2C_{2u}-C_{2d}$ ($u$ and $d$ standing for the up and the down quarks, respectively) with a 50\% uncertainty, with substantial improvement expected from the upcoming SoLID experiment \cite{Sou16}. However, the determination of $C_{2u}$ and $2C_{2d}$ individually is limited by the orthogonal linear combination $C_{2u}+2C_{2d}$, which is presently known with several times less precision.  Measurements of NSD-PV in light molecule systems are complimentary to the ongoing scattering-based measurements because $^9$Be and $^{25}$Mg possess an unpaired neutron, meaning these nuclei are primarily sensitive to $C_{2n} \simeq -0.4C_{2u}+0.8C_{2d}$ \cite{CahKan77}. Combined with PVDIS/SoLID, a precision NSD-PV measurement in one of the systems considered here would represent the first experimental determination of $C_{2u}$ and $C_{2d}$. 

The third contribution, $\kappa_\text{hfs}$, originates in the nuclear-spin-independent weak interaction combined with the hyperfine interaction \cite{FlaKhr85-2}. In the single-particle approximation, it is given by \cite{FlaKhr85-2,flambaum1997anapole}
\begin{eqnarray}
  \kappa_{\text{hfs}} = -\frac{1}{3} Q_W \frac{\alpha \mu}{m_p r_0 A^{1/3}} \simeq 2.5 \times 10^{-4} A^{2/3} \mu, 
\end{eqnarray}
where $\mu$ is the magnetic moment of the nucleus in units of nuclear magneton and $Q_W$ is the nuclear weak charge. The hyperfine interaction scales like $A^{2/3}$, similar to the anapole interaction, but due to the small numerical prefactor it is strongly suppressed.

Equations (\ref{eq:ka}) and (\ref{eq:kax}) estimate $\kappa_\text{A}$ and $\kappa_\text{ax}$ respectively in the single-particle (i.e.\ valence nucleon) limit.  This model ignores nucleon-nucleon interactions (apart from the parity-violating effects), and is an especially rough approximation for nuclei with partially-filled shells.  In Section \ref{sec:ncsm} we use a more sophisticated no-core shell model (NCSM) \cite{Barrett2013}  to calculate the anapole moments and $\kappa_\text{ax}$ of the $^9$Be, $^{13}$C, $^{14,15}$N, and $^{25}$Mg nuclei.

We should note another NSD-PV effect produced by the (tensor-type) interaction between the electrons and the nuclear weak quadrupole moment. Measurements of these moments will allow the first determination of the quadrupole moments of the neutron distribution in nuclei and provide a test of the theory of nuclear forces with applications to nuclei and neutron stars \cite{flambaum2016enhancing,flambaum2017effect,lackenby2018weak}. As with other NSD-PV effects, the effect of the nuclear weak quadruple moment is expected to be enhanced in certain systems \cite{SkrPetTit19}.

Eq. (\ref{HNSD}) can be rewritten for the $^2 \Sigma_{1/2}$ and $^2 \Pi_{1/2}$ electronic states \cite{Fla85,demille2008using} as
\begin{eqnarray}\label{eq:effective hamiltonian}
H_{\text{NSD-PV}}^{\text{eff}}=\kappa W_{\text{PV}} \Bigl(\bm{\hat{n}}\times \bm{S}_{\text{eff}} \Bigr)\cdot \bm{I}/I,
\end{eqnarray}
where $\bm{\hat{n}}$ is the unit vector pointing from the heavier to the lighter nucleus along the internuclear axis, and $\bm{S}_{\text{eff}}$ is the effective spin of the valence electron. In order to precisely determine the effective coupling constant $\kappa$ from experiments, the parameter $W_{\text{PV}}$ needs to be known with high accuracy. This parameter depends on the electronic structure and is specific to the given atom or molecule and to the electronic state. It is defined by the matrix element between two different $\Ket\Omega $ states \cite{KudPetSkr14}, 
\begin{eqnarray}
W_{\text{PV}} \equiv \frac{G_{\mathrm{F}}}{\sqrt{2}}\bra {+\tfrac{1}{2}}\rho
(\bm{r}){\alpha _{+}}\ket{-\tfrac{1}{2} } \ ,
\label{eqwa}
\end{eqnarray}
\noindent with
\begin{eqnarray}
\alpha_{+}=\alpha _{x}+i\alpha _{y}=\left(
\begin{array}{cc}
{0} & \sigma _{x} \\
\sigma _{x} & {0} \\
\end{array}
\right) +i\left(
\begin{array}{cc}
{0} & \sigma _{y} \\
\sigma _{y} & {0} \\
\end{array}
\right),
\end{eqnarray}
where $\sigma_x$ and $\sigma_y$ are the Pauli matrices and $\rho(\bm{r})$ is the nuclear density distribution function, which is assumed to have a Gaussian shape. $W_{\text{PV}}$ cannot be measured and has to be provided by sophisticated molecular calculations.

We use the relativistic coupled-cluster approach to determine the $W_{\text{PV}}$ coupling constants of the BeNC, BeCN, MgNC, and MgCN molecules with the highest possible accuracy; these results are presented in Section \ref{sec4}. This approach is considered to be the most powerful and accurate method for computational investigation of atomic and molecular properties. In the context of the NSD-PV it was previously applied to RaF \cite{KudPetSkr14}, HgH \cite{GedSkrBor18}, and BaF \cite{hao2018nuclear}. An advantage of this method is the possibility of setting uncertainty estimates on the obtained results, which we also do in the present work. To the best of our knowledge, no prior numerical investigations of the sensitivity of the above systems to the NSD-PV effects are available.

\section{No-core shell model nuclear calculations}\label{sec:ncsm}

In the NCSM, nuclei are considered to be systems of $A$ nonrelativistic point-like nucleons interacting via realistic two- and three-body interactions. Each nucleon is an active degree of freedom and the translational invariance of observables, the angular momentum, and the parity of the nucleus are conserved. The many-body wave function is expanded over a basis of antisymmetric $A$-nucleon harmonic oscillator (HO) states. The basis contains up to $N_{\text{max}}$ HO excitations above the lowest-possible Pauli configuration and depends on an additional parameter $\Omega$, the frequency of the HO well.

The only input for the present NCSM calculations was the Hamiltonian from Ref.~\cite{PhysRevC.101.014318} consisting of chiral nucleon-nucleon (NN) interaction obtained at the fourth order of chiral perturbation expansion (N$^3$LO) ~\cite{Entem2003} and chiral three-nucleon (3N) interaction at the N$^2$LO order denoted NN N$^3$LO + 3N(lnl). For a more efficient convergence, the Hamiltonian was renormalized by the Similarity-Renormalization-Group (SRG) unitary transformation~\cite{PhysRevC.75.061001, PhysRevLett.103.082501} with the evolution parameter $\lambda_{\text{SRG}}{=}2$ fm$^{-1}$. For $^9$Be, the largest basis space we were able to reach was $N_{\text{max}}{=}9$, while for the other p-shell nuclei we calculated up to $N_{\text{max}}{=}7$ using the importance truncation~\cite{PhysRevLett.99.092501,PhysRevC.79.064324} for $N_{\text{max}}{=}7$. The $^{25}$Mg is on the borderline of NCSM applicability. Only calculations up to $N_{\text{max}}{=}3$ were performed using importance truncation for $N_{\text{max}}{=}3$. The m-scheme dimensions of the largest basis spaces were of the order of $10^8$. The HO frequency of $\hbar\Omega{=}20$ MeV, optimised in Ref. \cite{PhysRevC.101.014318} was used.

The natural (i.e., ground-state) parity eigenstates are obtained in the even $N_{\text{max}}$ spaces while the unnatural parity eigenstates in the odd $N_{\text{max}}$ spaces. The parity non-conserving (PNC) NN interaction admixes the unnatural parity states in the ground state,
\begin{eqnarray}\label{gswf}
  |\psi_{\text{gs}}\; I \rangle &=& |\psi_{\text{gs}}\; I^\pi \rangle + \sum_j  |\psi_j \; I^{-\pi}\rangle \\ \nonumber
                                 &\times& \frac{1}{E_{\text{gs}}-E_j} \langle \psi_j \; I^{-\pi}| V_{\text{NN}}^{\text{PNC}}|\psi_{\text{gs}} \; I^\pi \rangle \; ,
\end{eqnarray}  
which then gives rise to the anapole moment. We used the Desplanques, Donoghue, and Holstein (DDH) PNC NN interaction of Ref.~\cite{DDH80} with their recommended parameter values except for the $f_\pi\equiv h_\pi^1{=}2.6\times10^{-7}$ taken from Ref.~\cite{BlythFryFom18}. In NCSM, when the $|\psi_{\text{gs}}\; I^\pi \rangle$ is calculated in $N_{\text{max}}$ space, the corresponding unnatural parity states appearing in Eq. (\ref{gswf}) are obtained in $N_{\text{max}}{+}1$ space. It is not necessary to compute many excited unnatural parity states as Eq.~(\ref{gswf}) suggests. Rather, the wave function $|\psi_{\text{gs}}\; I \rangle$ is obtained by solving the Schr\"{o}dinger equation with an inhomogeneous term
\begin{equation}\label{inhomeq}
  (E_{\text{gs}}-H)  |\psi_{\text{gs}}\; I \rangle =  V_{\text{NN}}^{\text{PNC}}|\psi_{\text{gs}} \; I^\pi \rangle \; .
\end{equation}
To invert this equation, we apply the Lanczos algorithm~\cite{Haydock_1974,Marchisio2003,STETCU2008168}.

In the presented calculations, we use the spin part of the anapole operator
\begin{equation}\label{as}
  \bm{a}_s=\frac{\pi e}{m} \sum_{i=1}^A \mu_i (\bm{r}_i\times\bm{\sigma}_i) \; ,
\end{equation}
which gives the dominant contribution to the anapole moment~\cite{PhysRevC.56.1641}. In Eq.~(\ref{as}), $m$ is the nucleon mass and $\mu_i$ is the nucleon magnetic moment in units of nuclear magneton, i.e., $\mu_i{=}\mu_p (1/2{+}t_{z,i})+\mu_n(1/2{-}t_{z,i})$ with $t_{z,i}{=}1/2$ ($-1/2$) for proton (neutron). The relationship between $\kappa_\text{A}$ and $a_s$ is given by
\begin{equation}\label{kappa_A_a_s}
\kappa_\text{A} = \frac{\sqrt2 e}{G_F} a_s,
\end{equation}
with 
\begin{equation}\label{as2}
a_s = \langle \psi_{\text{gs}} \; I \;  I_z{=}I | a^{(1)}_{s, 0} | \psi_{\text{gs}} \; I \;  I_z{=}I \rangle.
\end{equation}

Using Eqs.~(\ref{gswf}), (\ref{as}), (\ref{kappa_A_a_s}), and (\ref{as2})   we calculate the anapole moment similarly to Ref.~\cite{PhysRevC.60.025501} and find for the dimensionless coupling constant $\kappa_\text{A}$
\begin{eqnarray}\label{kappaA}
  \kappa_A &=& -i 4\pi \frac{e^2}{G_F}\frac{\hbar}{m c}\frac{(I I 1 0 | I I)}{\sqrt{2I+1}} \\ \nonumber
                      &\times& \sum_j \langle \psi_{\text{gs}} \; I^\pi||\sqrt{4\pi/3} \sum_{i=1}^A \mu_i r_i[Y_1(\hat{r}_i)\sigma_i]^{(1)} ||\psi_j \; I^{-\pi}\rangle \\ \nonumber
                                 &\times& \frac{1}{E_{\text{gs}}-E_j} \langle \psi_j \; I^{-\pi}| V_{\text{NN}}^{\text{PNC}}|\psi_{\text{gs}} \; I^\pi \rangle \; ,
 \end{eqnarray}  
 where $(I I 1 0 | I I){=}I/\sqrt{I(I+1)}$.

We have also performed NCSM calculations for the matrix elements of the spin operators that serve as input for the calculation of the coupling constant $\kappa_\text{ax}{=} -2 C_{2p} \langle s_{p,z} \rangle - 2 C_{2n} \langle s_{n,z} \rangle {\simeq}-0.1 \langle s_{p,z}\rangle{+}0.1 \langle s_{n,z}\rangle$.
 The spin operator matrix elements are defined as
 \begin{eqnarray}
 \langle s_{\nu,z} \rangle{\equiv}\langle \psi_{\text{gs}} \; I^\pi I_z{=}I|s_{\nu,z}|\psi_{\text{gs}} \; I^\pi I_z{=}I\rangle, 
\end{eqnarray}
with $\nu{=}p,n$.

\begin{figure}[t]
\includegraphics[scale=0.35]{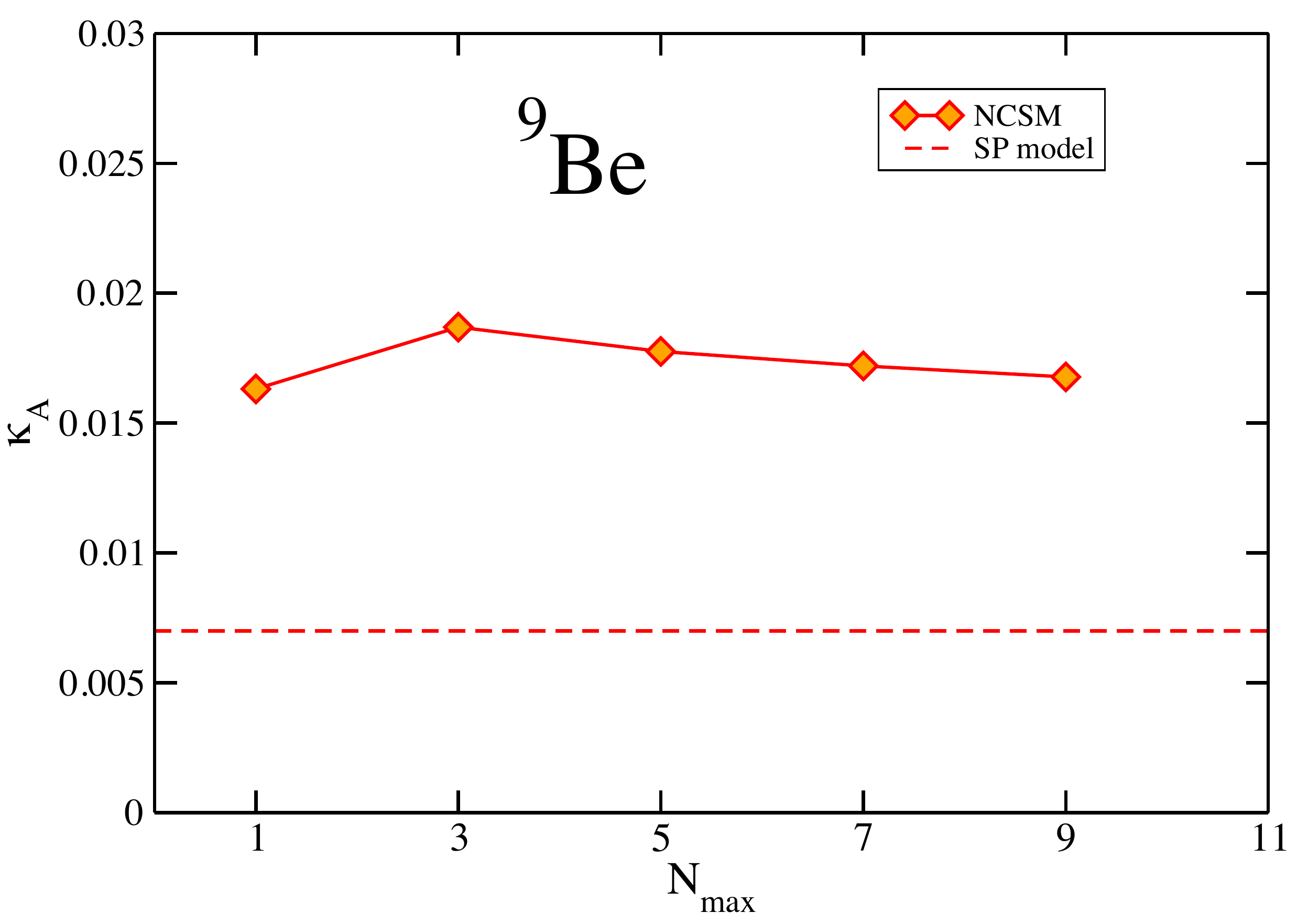}
\caption{Dependence of the anapole-moment coupling constant $\kappa_\text{A}$ for $^9$Be on the size of the NCSM basis characterized by $N_{\text{max}}$. The dashed line represents $\kappa_\text{A}$ obtained in the single-particle model.}
\label{fig9Be}
\end{figure}

Our results for the anapole-moment coupling constants $\kappa_\text{A}$  and $\kappa_\text{ax}$ in $^9$Be, $^{13}$C, $^{14,15}$N, and $^{25}$Mg are summarised in Table~\ref{tabNCSMkappa}. Overall, the basis size convergence of the results is quite reasonable, as shown in Fig.~\ref{fig9Be} presenting the dependence of $\kappa_\text{A}$ of $^9$Be on the NCSM basis size characterised by $N_{\text{max}}$. We can thus evaluate the uncertainties due to the basis size convergence at about 10\% (25\% for $^{25}$Mg). The other sources of uncertainty are renormalization and incompleteness of the transition operators and the uncertainties due to the description of nuclear and the parity-violating forces.

In Table~\ref{tabNCSMkappa}, we also present NCSM results for magnetic moments, where we can compare our results with experimental values. Overall, we find a qualitative agreement with experiment with some underestimation of absolute values. This is not surprising, since the present calculations included only the one-body M1 operator. It is well established that two-body currents contribute non-negligibly to M1 matrix elements in light nuclei~\cite{Pastore:2015dho}. While the dominant sources of uncertainty are different for the calculated dipole moments and the NSD-PV parameters, we can still use the deviation of the former from experiment as a rough estimate of the accuracy of the calculations of the latter.

Table~\ref{tabNCSMkappa} also contains the single-particle model estimates of the different contributions to the NSD-PV constant  $\kappa = \kappa_\text{A} + \kappa_{\text{ax}} + \kappa_{\text{hfs}}$ obtained using equations (2-5) for the nuclei in the molecules considered in the present work. Note that the $^{14}$N nucleus contains a valence proton and a valence neutron, both in the $p_{1/2}$ orbital with $K=1$.  The nuclear magnetic moment $\mu  \simeq 0.404$  is given, to a good accuracy, by the sum of the magnetic moments of $^{13}$C (with valence $p_{1/2}$ neutron) and $^{15}$N (with valence $p_{1/2}$ proton). Therefore, we took the sum of the valence proton and neutron contributions for the other constants.

The NCSM $\kappa_\text{A}$ results are higher in absolute values than the single-particle model ones by a factor of 2--4, except for $^{14}$N. The largest differences are found in the mid-shell nuclei $^9$Be, $^{13}$C and $^{25}$Mg, for which the single-particle model has limited applicability. The $^{14}$N anapole moment is proportional to the sum of the $^{15}$N and $^{13}$C anapole moments that have opposite signs and consequently it is particularly sensitive to the $V_{\text{NN}}^{\text{PNC}}$ parametrisation and the other computational details. The NCSM  $\kappa_\text{ax}$ results are close to the single-particle model for $^{13}$C and $^{15}$N while they differ more substantially for the mid-shell $^9$Be and $^{25}$Mg. For $^{14}$N, the $\kappa_\text{ax}{\simeq} 0$ as $\langle s_{p,z}\rangle{\simeq}-\langle s_{n,z}\rangle$.

\begin{table}[t]
\caption{ Magnetic moments (in units of nuclear magneton) \cite{dickinson1949magnetic,royden1954measurement,fuller1976nuclear,baldeschwieler1962double,alder1951magnetic,stone2005table}, anapole-moment coupling constants, spin operator matrix elements, and $\kappa_{\text{ax}}$ coupling constants for  $^9$Be, $^{13}$C, $^{14,15}$N and $^{25}$Mg obtained within NCSM. The results obtained using the single-particle model are also shown, along with the valence particle (V.p.) and the valence orbital (V.o) for each nucleus.}
\label{tabNCSMkappa}
\begin{center}
\begin{ruledtabular}
  \begin{tabular}{lccccc}
       & $^9$Be    &  $^{13}$C  &  $^{14}$N   &  $^{15}$N  &  $^{25}$Mg  \\ \hline
$I^\pi$ &  $3/2^-$ &  $1/2^-$   &    $1^+$     &   $1/2^-$ &    $5/2^+$   \\
$\mu^{\text{exp.}}$            & -1.177$^a$  &  0.702$^b$    & 0.404$^c$   &  -0.283$^d$  &    -0.855$^e$    \\ 
            & \multicolumn{5}{c}{\underline{NCSM calculations}} \\   
$\mu$                            & -1.05  &  0.44   & 0.37 &  -0.25 &    -0.50   \\
$\kappa_\text{A}$                   & 0.016& -0.028 & 0.036  & 0.088 &  0.035 \\                   
$\langle s_{p,z} \rangle$ &  0.009    &   -0.049   &  -0.183      &  -0.148     &    0.06       \\
$\langle s_{n,z} \rangle$ &  0.360     &   -0.141   &  -0.1815    &    0.004     &    0.30       \\
$\kappa_\text{ax}$                 & 0.035  & -0.019 &  0.0002 &  0.015 & 0.024   \\
$\kappa$                 & 0.050  & -0.046 &  0.037 &  0.103 & 0.057   \\
  & \multicolumn{5}{c}{\underline{Single-particle model calculations}} \\
   V. p.&$n$&$n$&$n$, $p$&$p$&$n$\\
  V. o.&$p_{3/2}$&$p_{1/2}$&$p_{1/2}$&$p_{1/2}$&$d_{5/2}$\\
  $K$             &   -2    &1  &1 &1&-3\\ 
     $\kappa_\text{A}$             &   0.007    &-0.007  &0.035 &0.044&0.014\\
   $\kappa_{\text{ax}}$             &   0.050    &-0.017 &0.0  &0.017&0.050\\
   $\kappa_{\text{hfs}}$             &  -0.001   &0.001 &0.0006  &-0.0004&-0.002\\
   $\kappa$             &  0.056   &-0.023 &0.036 &0.060&0.062
  \end{tabular}
\end{ruledtabular}
\end{center} 
\begin{tablenotes}                                                                  \footnotesize
\item $^a$ Refs. \cite{dickinson1949magnetic,stone2005table}; $^b$ Refs. \cite{royden1954measurement,stone2005table}; $^c$ Refs. \cite{fuller1976nuclear,stone2005table}; $^d$ Refs. \cite{baldeschwieler1962double,stone2005table}; $^e$ Refs. \cite{alder1951magnetic,stone2005table}.
\end{tablenotes}
\end{table}

The results obtained within the single-particle model predict that the $Z^0$-boson exchange constant $\kappa_{\text{ax}}$ dominates for the light nuclei containing a valence neutron, that is $^{25}$Mg, $^{13}$C, and $^9$Be are significantly more sensitive to $\kappa_{\text{ax}}$, while in the $^{14}$N and $^{15}$N nuclei the anapole-moment effect dominates. However, a different picture emerges from the NCSM calculations: $\kappa_{\text{ax}}$ still dominates in  $^9$Be, while $^{14}$N and $^{15}$N are more sensitive to the anapole moments, and $^{25}$Mg and $^{13}$C have roughly the same sensitivities to the two effects. Furthermore, within the single-particle model, the total NSD-PV effect is roughly equivalent in $^9$Be, $^{15}$N, and $^{25}$Mg, while the NCSM calculations predict a significantly larger total $\kappa$ in $^{15}$N. This difference highlights the importance of using sophisticated computational methods beyond the single-particle model. 

The presented NCSM calculations of the anapole moments of light nuclei can be improved in several ways. First, higher-order terms in the anapole-moment operator including two-body current contributions~\cite{PhysRevC.65.045502} should be included. Second, both the anapole-moment operator and the PNC NN interaction should be SRG renormalized consistently with the nuclear chiral Hamiltonian. The technical capability to do this in the NCSM has been developed~\cite{Gysbers2019} and the renormalization calculations are underway. We anticipate that the SRG renormalization will reduce the $\kappa_\text{A}$ in absolute value. At the same time, we note that the calculated $E_{\text{gs}}$ energies in Eqs.~(\ref{gswf}) and (\ref{kappaA}) could be phenomenologically corrected so that the excitation energies of the lowest unnatural parity states match experimental values. This would enhance the absolute values of $\kappa_\text{A}$. This correction was not applied here to compensate to some extent for the lack of the SRG renormalization. One should also explore the sensitivity of the results to the form of the PNC interaction by applying the recently derived chiral PV nuclear forces~\cite{deVries:2020iea}. Given the good basis size convergence of both $\kappa_\text{A}$ and $\kappa_{\text{ax}}$ found in the present calculations, we can be optimistic that uncertainties of nuclear calculations for light nuclei can be reduced to $\sim10\%$ once the above improvements are implemented. The NCSM used here, as well as the valence-space in-medium similarity renormalization group (VS IM-SRG), are suitable treatments for nuclei with partially filled shells, but current state-of-the-art of multi-shell calculations (required for anapole moments) is  limited to systems up to about Mg \cite{miyagi2020ab}.

Many heavy nuclei are currently targets of NSD-PV measurements. While the single-particle model is appropriate for a nucleus with a single nucleon or hole outside a closed shell, such as $^{133}$Cs \cite{WooBenCho97} or $^{137}$Ba \cite{AltAmmCah18-1}, our work  highlights the importance of extending our computational framework to  model heavy nuclei with partially filled shells, such as $^{163}$Dy \cite{LeeGouAnt14}, $^{171}$Yb \cite{AntFabBou17}, and $^{212}$Fr \cite{AubBehCol13}.

\section{Electronic structure calculations of the $W_{\text{PV}}$ coupling constants. \label{sec4}}

The main factors that determine the quality of a calculation of any molecular property are the treatment of relativity and electron correlation and the choice or the basis set. For high-accuracy determination of the $W_{\text{PV}}$ parameters, relativistic effects should be incorporated into the calculations and correlation should be treated within a state-of-the-art approach. Furthermore, high-quality basis sets should be used to provide a good description of the electronic wavefunctions, especially in the region near the nuclei.

In this work, the calculations are performed using the developer's version of the DIRAC program package \cite{DIRAC15}, within the Dirac-Coulomb Hamiltonian,
\begin{eqnarray}
H_{0} = \sum\limits_{i} \Bigl[c\boldsymbol\alpha_i \cdot \bm{p}_{i} + \beta_i m_e c^2 + V(r_i) \Bigr] + \sum\limits_{i<j} \frac{e^2}{r_{ij}}.
\label{H0}
\end{eqnarray}
The Coulomb potential $V(r_i)$ takes into account the finite size of the nuclei, modelled by Gaussian charge distributions \cite{VisDya97}.

In order to investigate the impact of treatment of the electron correlation on the calculated $W_{\text{PV}}$ parameters, we compare open-shell single determinant average-of-configuration Dirac-Hartree-Fock \cite{Thy01} and the single-reference relativistic coupled-cluster method with single and double excitations (CCSD) and augmented with perturbative triple excitations (CCSD(T)) \cite{VisLeeDya96}.

We employ Dyall's relativistic basis sets \cite{Dya09,Dya16} of varying quality to examine the basis set size effects on the calculated $W_{\text{PV}}$ parameters. We also augment these basis sets manually with specific functions needed for improving the quality of the calculations (in particular for description of the nuclear region).

The first step in our study is to establish the equilibrium geometries of the molecules (only that of MgNC is available from experiment \cite{AndZiu94}). We thus perform molecular geometry optimizations, using the relativistic CCSD(T) approach and the dyall.v4z basis sets, augmented with one extra diffuse function for each symmetry (s-aug-dyall.v4z). The calculated equilibrium bond lengths together with the results from the previous studies for BeNC, BeCN, MgNC, and MgNC are summarized in Table \ref{Retab}. 

\begin{table}[t]
\caption{Equilibrium bond lengths of BeNC, BeCN, MgNC, and MgCN (\AA).
}
\begin{ruledtabular}                                                             
\begin{tabular}{lllll}
& $R_1$ & $R_2$ & Method & Ref.\\\hline
BeNC &1.528 &1.181 &CCSD(T) & This work\\
     &1.57  &1.19  &DFT & \cite{LanAnd97}\\

BeCN &1.668 &1.164 &CCSD(T) & This work\\
     &1.69  &1.17  & DFT & \cite{LanAnd97}\\

MgNC &1.931 &1.179 &CCSD(T) & This work\\
     &1.925 &1.169 &Exp.    & \cite{AndZiu94}\\
     &1.947 &1.181 &CCSD(T) & \cite{woon1996ab}\\

MgCN &2.069 &1.166 &CCSD(T) & This work\\
     &2.074 &1.168 &CCSD(T) & \cite{woon1996ab}\\
\end{tabular}
\end{ruledtabular}
\begin{tablenotes}                                                  \footnotesize
\centering
\item[\emph{a}]{
$R_1$: equilibrium distance between the first and the second atoms.
}
\item[\emph{b}]{
$R_2$: equilibrium distance between the second and the third atoms.
}
\end{tablenotes}
\label{Retab}
\end{table}

Our results are in good agreement with the experimental geometry of the MgNC molecule, and with the earlier CCSD(T) results (obtained with somewhat smaller basis sets), lending credence to our predictions for the rest of the molecules, where experimental data are lacking.  For BeNC and BeCN, the only previous study was performed within the DFT approach \cite{LanAnd97} and those results have larger discrepancies with the current predictions, in particular for $R_1$. It is worth mentioning that equilibrium bond lengths between the first and the second atoms ($R_1$) in isocyanides are shorter than that in cyanides due to the higher electronegativity of nitrogen compared to carbon.

The calculations of the $W_{\text{PV}}$ parameters are carried out within the framework of the finite-field (FF)  approach \cite{Mon77,Thyssen2000}, where the NSD-PV interaction is added as a perturbation. In this approach, the total Hamiltonian includes the usual unperturbed term defined in Eq. (\ref{H0}) and a perturbative term arising from Eq. (\ref{eqwa}), 
\begin{eqnarray}
H(\lambda_N)=H_{0}+ \frac{G_{\mathrm{F}}}{\sqrt{2}}\lambda_N \rho (\bm{r}
_{N}){\alpha _{+}},
\label{Hlamda}
\end{eqnarray}
\noindent where $\lambda_N$ is a small perturbation parameter describing the strength of the effective NSD-PV effect for the nucleus $N$ and $\rho(\bm{r}_N)$ is the nuclear density distribution function of the corresponding nucleus. For triatomic molecules there are three $W_{\text{PV}}$ values (one for each nucleus). Since the perturbation parameters $\lambda_N$ are small, the total energy can be expanded around $\lambda_N=0$. We then obtain
\begin{eqnarray}
E(\lambda_N)=E(0)+ \lambda_{N}\frac{\partial E(\lambda_N)}{\partial \lambda_N}\bigg|_{\lambda_N \rightarrow 0}+...
\label{Elamda}
\end{eqnarray}

Combining Eq. (\ref{Elamda}) with Eqs. (\ref{eqwa}) and (\ref{Hlamda}) and invoking the Hellmann-Feynman theorem we get
\begin{eqnarray}
W_{\text{PV}}(N) \equiv \frac{\partial E(\lambda_N )}{\partial \lambda_N }\bigg|_{\lambda_N \rightarrow 0}.
\end{eqnarray}

In order to obtain good linearity in the energy dependence on $\lambda_N$, so that the higher-order terms can be ignored while not losing the total energy shift within the precision of the calculation, selection of the optimal field strength is important.  
Since the effects tend to be small for light nuclei, we use  fields of $10^{-5}$ a.u. for Be and C and smaller fields of $10^{-6}$ a.u. for Mg and N. Moreover, we require the energy convergence of the coupled-cluster iterations to be $10^{-12}$ a.u. In practice, we compute the total energies using two field strengths, which are taken symmetrically with respect to zero, and employ the 2-point formula to obtain the $W_{\text{PV}}$ parameters. This is done separately for each nucleus.

\begin{table}[t]
\caption{$W_{\text{PV}}$ parameters (Hz), calculated at the equilibrium geometry and using the optimised basis sets (see text for further detail).
}
\begin{ruledtabular}                                                             
\begin{tabular}{llccc}
Molecule&Nucleus&DHF&CCSD&CCSD(T)\\\hline
               
               &Be      &0.402   &0.498  &0.498\\          
BeNC           &N       &0.169   &0.331  &0.340\\
               &C       &0.006   &0.005  &0.004\\\hline

               &Be      &0.440   &0.542  &0.540\\  
BeCN           &C       &0.124   &0.269  &0.278\\
               &N       &0.017   &0.029  &0.030\\\hline

               &Mg      &4.016   &5.298  &5.297\\
MgNC           &N       &0.191   &0.443  &0.454\\
               &C       &0.012   &0.015  &0.014\\\hline
               
               &Mg      &4.022   &5.310  &5.310\\
MgNC$^\dagger$ &N       &0.191   &0.457  &0.469\\
               &C       &0.012   &0.015  &0.014\\\hline

               &Mg      &4.099   &5.430  &5.419\\  
MgCN           &C       &0.143   &0.363  &0.375\\
               &N       &0.029   &0.063  &0.064\\
               
\end{tabular}
\end{ruledtabular}
\begin{tablenotes}                                                                                  \footnotesize
\centering
\item[\emph{a}]{
$^\dagger$ Results obtained using experimental geometry, Ref. \cite{AndZiu94}.
}
\end{tablenotes}
\label{WAtab}
\end{table}

Using the optimised molecular geometries and the finite field approach, we performed calculations of the $W_{\text{PV}}$ parameters. The basis set tests we have carried out show that in terms of basis set quality, the results tend to be converged at the v4z level. Switching to the core-valence, cv4z basis sets, which include two extra d-type tight (high-exponent) functions and one extra f-type tight function changes the calculated $W_{\text{PV}}$ of the metal atoms by less than 0.001\%. Furthermore,  addition of diffuse functions (using singly augmented v4z basis, s-aug-v4z) has very little effect on the results, less than 0.5\,\% in all cases. On the other hand, we found that the calculated constants are sensitive to the presence of tight s and p functions, as was also observed in our previous works \cite{hao2018nuclear,BorIliDzu12-1,BorIliDzu13}. This is due to the fact that such functions improve the description of the electronic wave function in the nuclear region. We have added tight p and s functions (designated as ts, tp) to the standard v4z basis sets until we obtained convergence of the calculated $W_{\text{PV}}$ constants. The final basis sets are v4z+4tp for Be and Mg and v4z+3ts+4tp for C and N. The $W_{\text{PV}}$ parameters calculated at the equilibrium bond lengths and using the optimised basis sets are presented in Table~\ref{WAtab}. For the coupled-cluster calculations, virtual space cutoff of 200 a.u. was used.

\begin{figure}[t]
\centering
\includegraphics[scale=0.99,width=0.99\linewidth]{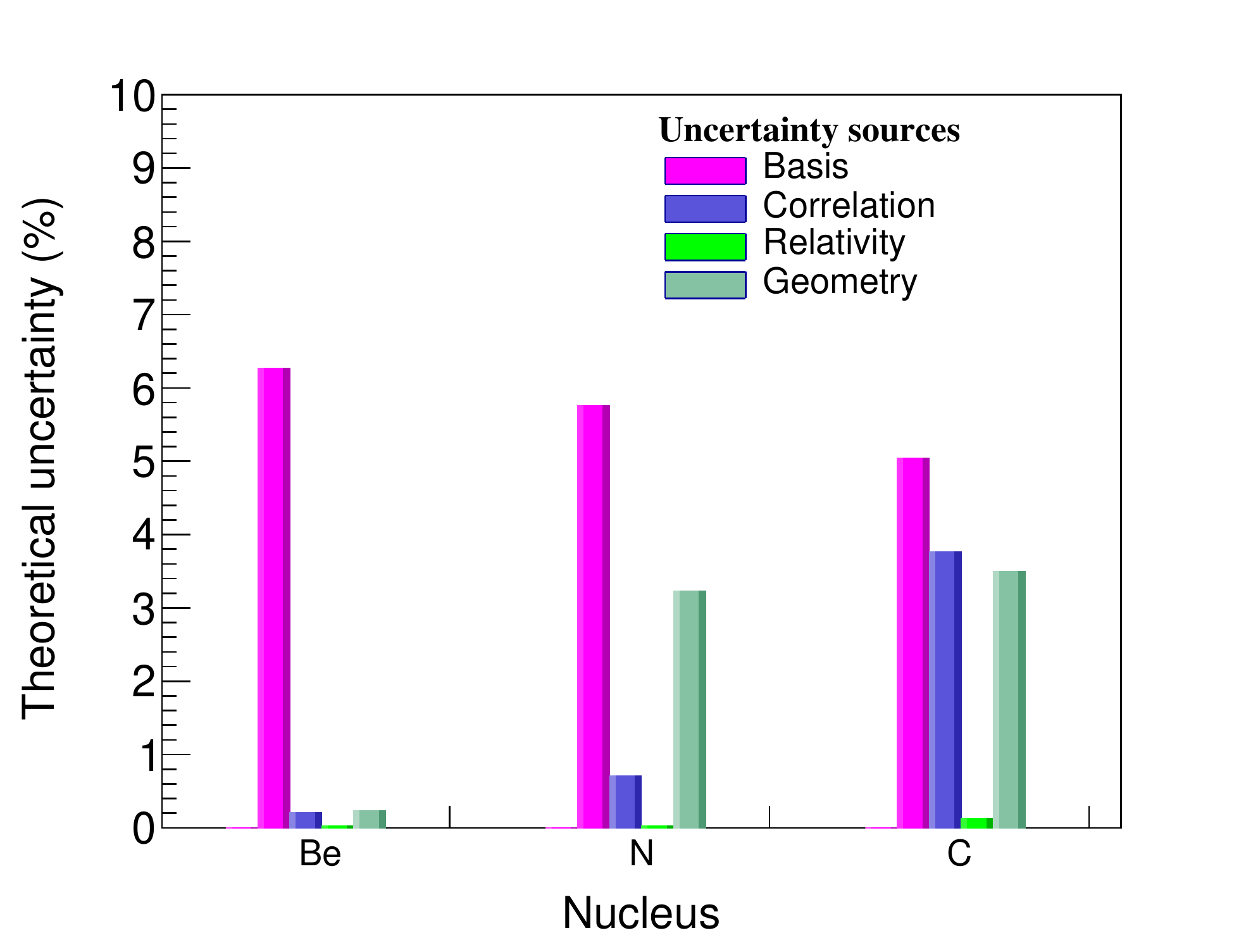}
\includegraphics[scale=0.99,width=0.99\linewidth]{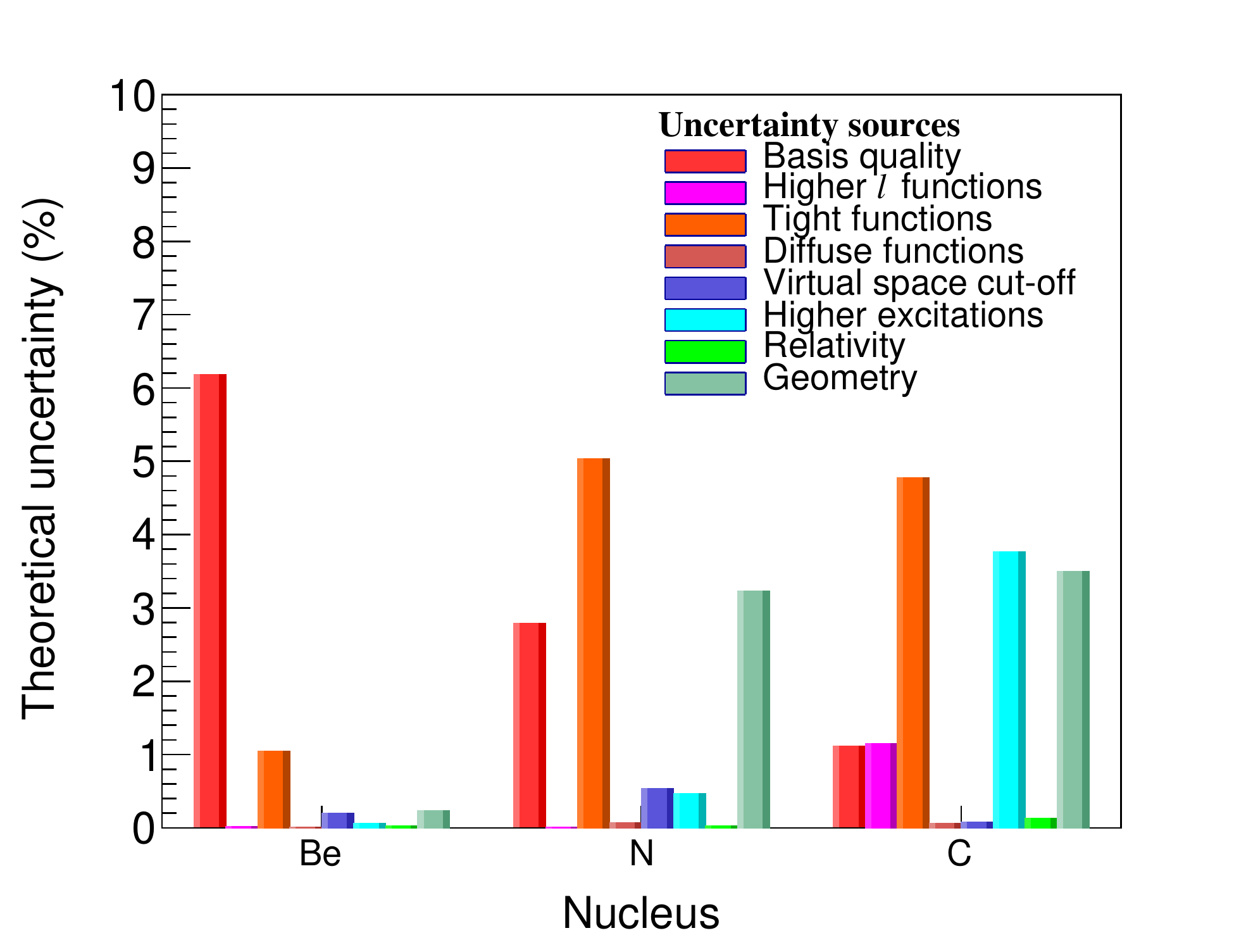}\\
\caption{Theoretical uncertainties in the calculated $W_{PV}$ parameters of Be, N, and C in BeNC, given in $\%$ with respect to the final result. The lower panel presents a detailed breakdown of various sources of uncertainty.}
\label{edmpt}
\end{figure}

For the metal atoms, the electron correlation contributes approximately $20\%$ to the $W_{\text{PV}}$ values, while the effect of the triple excitations is negligible. As expected, the $W_{\text{PV}}$ values for Mg are larger than those for Be, in keeping with the predicted scaling of the effect roughly as $Z^{2}$ for light elements \cite{FlaKhr85-1}. Furthermore, the ligand structure (cyanide vs. isocyanide compounds) has negligible effect on the $W_{\text{PV}}$ of the metal atoms; the results obtained here are also in good agreement with those for the diatomic BeF and MgF compounds investigated in our earlier work \cite{hao2018nuclear}.  From Eq. 7, the $W_{\text{PV}}$ parameter depends on the overlap of the wavefunction of the unpaired electron with the nucleus. Our results are consistent with the unpaired electron being largely centred on the metal atom in all the species considered here. This property is also an identifying feature of laser coolable molecules:  an unpaired electron that is highly centred on a single atom in both the ground and excited electronic states allows for optical cycling with minimal vibrational branching \cite{norrgard2019nuclear}.  The observation that $W_{\text{PV}}$ is significantly reduced for the third atom in all cases is again the result of the unpaired electron being centred on the metal atom.  For example, in BeCN, $W_{\text{PV}}$(Be) (about  0.5 Hz) is nearly 100 times larger than $W_{\text{PV}}$(N) (about 0.008  Hz), despite N being the heaviest nucleus in this molecule.
 
The correlation effects play a larger role for the  $W_{\text{PV}}$ parameters of carbon and nitrogen, and contribute up to 50$\%$ of their total values; furthermore, the contribution of the triple excitations is on the order of 2-10$\%$. This is due to the fact that the molecular bonding effects become important for the sensitivity of the ligand to the NSD-PV phenomena, and high-quality treatment of electron correlation is needed to capture these effects.
 
The  $W_{\text{PV}}$ values are used to interpret the experimental results. Therefore, it is important to estimate their uncertainty, which we do through a comprehensive investigation of the effect of the various computational parameters on the calculated  $W_{\text{PV}}$. The main remaining sources of uncertainty (for a relativistic CCSD(T) calculations with the optimised basis sets) are the remaining basis set incompleteness, the neglect of the relativistic effects beyond the DC Hamiltonian, the neglect of the higher excitations (beyond perturbative triples) and the effect of the virtual space cutoff in the correlation treatment. We address each of these separately by investigating the BeNC molecule; the different contributions are summarised in Table~\ref{taberror} for each of the atoms.

The effect of the basis set quality is evaluated by taking the difference between the results obtained with the v4z and the v3z basis sets, on the assumption that the residual basis set incompleteness should be smaller than this difference. The effect of lacking tight high angular momentum functions is extracted from comparison of a v4z basis set calculation with a cv4z basis one. We then consider the effect of augmentation of the basis set by comparison with the s-aug-v4z for the diffuse functions. For the tight functions, we use as estimate the change in the result upon removing one tight s and one tight p function from the optimised basis set.  
 
In order to test the effect of the virtual space cut-off we performed the calculations with a higher cut-off of 340 a.u (which includes an additional set of virtual orbitals in relation to the original cut-off), and took the difference between these results and those obtained with a cut-off of 200 a.u. as the corresponding uncertainty. The contribution of perturbative triple excitations is quite small, in all the cases. To estimate the error due to the neglect of full triple and higher-order excitations, we follow the scheme introduced in Ref. \cite{hao2018nuclear}, and use the spread in the values obtained with different schemes of inclusion of the triple excitations, namely, we take twice the difference between the CCSD+T and CCSD-T values.
 
We use twice the size of the Gaunt contribution \cite{Gau29}, calculated at the SCF level, to estimate the uncertainties due to the neglect of the Breit interaction and the QED effects. This contribution is extremely small for the 3 atoms, due to their low atomic number.

The remaining source of error is due to the molecular geometry. For MgCN, BeCN, and BeNC we calculate the $W_{\text{PV}}$ values at optimised equilibrium geometries; such optimisation carries a certain error. We can evaluate the resulting uncertainty by comparing the results obtained with the optimised and the experimental geometries in MgNC (Table~\ref{WAtab}). We take the difference between these values multiplied by 1.5 as a conservative estimate of the uncertainty due to the possible errors in the molecular geometry. Assuming MgNC and the rest of the systems to be structurally similar, we use the same uncertainty in \% for the four molecules.

Since we are dealing with higher-order effects, we assume the different sources of error to be largely independent, and thus obtain the total errors using quadratic sum (Table \ref{taberror}). The total relative uncertainties are smallest for the metal and largest for the outer ligand. The weight of the contributions of the different sources of uncertainty is also different for the three atoms (Fig. 2). These differences  reflect the position of the atom in the molecule, and the effects of the electronic structure and bonding. For the three atoms, the contribution of the basis set effects to the uncertainty is the largest. However, while this is the clearly dominating contribution in case of Be, for the middle atom the uncertainty in the geometry also plays an important role. For carbon, these two sources of uncertainty are augmented by that stemming from the neglect of the higher-order excitations, due to the higher sensitivity of the outer ligand to correlation effects.

\begin{table}[t]
\caption{Summary of the main sources of theoretical uncertainty in the calculated $W_{\text{PV}}$ parameters of Be, C, N in BeNC, given in $\%$ with respect to the final result.}
\label{taberror}
\begin{tabular*}{\columnwidth}{l@{\extracolsep{\fill}}lccc}
\hline\hline
\multicolumn{2}{l}{Error source}         & Be      & N      & C    \\
\hline
\multicolumn{2}{l}{Basis quality}        & 6.18    & 2.79   & 1.11  \\       
\multicolumn{2}{l}{Higher $l$ functions} & 0.02    & 0.01   & 1.15  \\
\multicolumn{2}{l}{Basis augmentations}  &         &        &       \\
&Tight functions                         & 1.05    & 5.03   & 4.78  \\
&Diffuse functions                       & 0.01    & 0.07   & 0.06  \\
\multicolumn{2}{l}{Correlation}          &         &        &       \\
&Virtual space cut-off                   & 0.20    & 0.54   & 0.08  \\
&Residual triples and                    &         &        &       \\
&higher excitations                      & 0.06    & 0.47   & 3.77  \\
\multicolumn{2}{l}{Relativity}           & 0.03    & 0.03   & 0.13  \\

\multicolumn{2}{l}{Geometry}             & 0.35    & 4.84   & 5.25  \\
\multicolumn{2}{l}{\textbf{Total ($\%$)}}& 6.29    & 7.56   & 8.20  \\   
\multicolumn{2}{l}{\textbf{Total (Hz)}}  & 0.0313  & 0.0257 & 0.0003\\
\hline
\hline
\end{tabular*}
\end{table}

\begin{table}[t]
\caption{Recommended values of the $W_{\text{PV}}$ parameters (Hz) with corresponding uncertainties.
}
\begin{ruledtabular}                                                             
\begin{tabular}{l|cccc}
\multirow{2}{*}{\diagbox{Mol.}{Atom}}&\multicolumn{2}{c}{Atom 1}&Atom 2&Atom 3\\ 
&Be&Mg& & \\ \hline

BeNC              &0.50   &$-$      &0.34   &0.004 \\     
BeCN              &0.54   &$-$      &0.28   &0.030 \\ 
MgNC              &$-$      &5.3   &0.45   &0.014 \\ 
MgNC$^\dagger$    &$-$      &5.3   &0.47   &0.014 \\ 
MgCN              &$-$      &5.4   &0.37   &0.064 \\ 
Uncertainty ($\%$)&6.3     &4.9     & 7.6    &8.2   \\
\end{tabular}\label{tab:recommend}
\end{ruledtabular}
\begin{tablenotes}                                                                      
\footnotesize
\centering
\item[\emph{a}]{
$^\dagger$ Results obtained using experimental geometry \cite{AndZiu94}.
}
\end{tablenotes}
\label{WAerrtab}
\end{table}

We use the results of the investigation on BeNC to assign uncertainties on the $W_{\text{PV}}$ values of Be, N, and C in the rest of the systems. However, we see that both the absolute $W_{\text{PV}}$ and the uncertainties are determined by the position of the atom in the molecule, rather than by its atomic number. We thus base the uncertainty evaluation on the position: for example, to evaluate the uncertainty of the calculated $W_{\text{PV}}$ of N (the outer ligand) in BeCN, we use the estimate of the uncertainty of C in BeNC, where C had an outer position. The uncertainty for Mg was obtained by a separate study on MgNC.

Table~\ref{tab:recommend} contains the final recommended $W_{\text{PV}}$ obtained at the CCSD(T) level of theory, along with the corresponding uncertainties.

\section{Experimental Considerations}

Testing the Standard Model NSD-PV effects in light nuclei calculated in Sections \ref{sec:ncsm} and \ref{sec4} requires high sensitivity measurements of the effective NSD-PV Hamiltonian (Eq.\,(\ref{eq:effective hamiltonian})) matrix elements $iW$.
A shot noise-limited measurement of $iW$ has uncertainty $\delta W = 1/\tau\sqrt{\mathcal{R} N T}$, where $\tau$ is the interaction time of a single measurement, $\mathcal{R}$ is the repetition rate, $N$ the number of trapped molecules per measurement, and $T$ the total measurement time.  Ref.\,\cite{norrgard2019nuclear} details a Stark interference scheme \cite{demille2008using,AltAmmCah18-1,AltAmmCah18-2} for measuring NSD-PV in polyatomic molecules in an optical dipole trap.  There, it is expected that differential ac Stark shifts from the trapping laser limit $\tau\lesssim 2$\,ms.  Assuming $\mathcal{R}=10\,\text{s}^{-1}$ and $N=1000$ molecules, the expected statistical uncertainty is  $\delta W \sim 2\pi\times1$ $\text{Hz}/\sqrt{\text{Hz}}$. 

Here we consider an alternative measurement scheme based on Stark interference of laser-cooled molecules in free flight with the potential for even greater sensitivity.  We begin by reviewing in brief the Stark interference method to highlight the increased sensitivity when measuring NSD-PV between nearly-degenerate opposite-parity states.  Next, we use  $^{25}$MgNC as an example to show that the first excited bending mode of linear polyatomic molecules may have several parity violation-sensitive state pairs which are tuneable to near-degeneracy with low ($\lesssim\,100\,$G) magnetic fields.  We conclude this section by estimating the free-flight measurement scheme's sensitivity to NSD-PV and consider likely leading broadening mechanisms and systematic effects.

\subsection{Stark Interference}
Because the effective NSD-PV Hamiltonian Eq.\,(\ref{eq:effective hamiltonian}) is a pseudoscalar interaction, it connects opposite parity states with the same total angular momentum projection $m_F$.
The NSD-PV signal may be measured by the Stark interference between two such states \cite{Nguyen1997,demille2008using,Tsigutkin2009,Cahn2014}.  
Denoting the time-dependent probability amplitudes of these states $c_\pm(t)$, and assuming an initial state $c_-(0)$\,=\,1, $c_+(0)$\,=\,0, the system evolves in the presence of an oscillating electric field $\mathcal{E} = E_0 \cos(\omega_E t) \hat{\mathbf{z}}$.
 The effective Hamiltonian $H_\pm^{\text{eff}}$ for the two level system is \cite{Cahn2014}
\begin{equation}\label{eq:tls}
  H_\pm^{\text{eff}} = \left( \begin{matrix}
                              \Delta & d\,E_0\,\text{cos}(\omega_E\,t) + i\,W \\
                              d\,E_0\,\text{cos}(\omega_E\,t) - i\,W & -\alpha^\prime\,{E_0}^2\cos^2(\omega_E\,t)/2
                            \end{matrix} \right).
\end{equation}
Here $\Delta$ is the energy difference of the opposite-parity states, $d$ is the transition dipole moment, $\alpha^\prime$ is the differential polarizability of the two states, and $iW$ is the matrix element of Eq.\,(\ref{eq:effective hamiltonian}) we wish to measure.   In the limit where  $W \ll d\,E_0, \Delta \ll \omega_E$, and  $\alpha^\prime = 0$, the PV signal $S$\,=\,$\vert c_+(t) \vert^2 $ is
\begin{equation}\label{eq:probability amplitude}
  S \simeq 4 \left[2\frac{W}{\Delta}\frac{d\,E_0}{\omega_E}+\left(\frac{d\,E_0}{\omega_E}\right)^2\right] \text{sin}^2 \left(\frac{\Delta\,t}{2}\right).
\end{equation}

The first term in square brackets of Eq.\,\ref{eq:probability amplitude} is the combined PV and electric dipole transition probability.  This term changes sign under a reversal of either $\mathcal{E}$ or $\Delta$.  The PV matrix element $iW$ may be extracted through an asymmetry measurement \cite{demille2008using}
\begin{equation}\label{eq:asymmetry}
  \mathcal{A}= \frac{S(+E_0)-S(-E_0)}{S(+E_0)+S(-E_0)} = 2 \frac{W}{\Delta} \frac{\omega_E}{d\,E_0}+ \ldots,
\end{equation}
where the ellipsis denotes higher-order terms in $W/\Delta$.

\begin{figure}[t]
   \centering
   \includegraphics[width= \columnwidth]{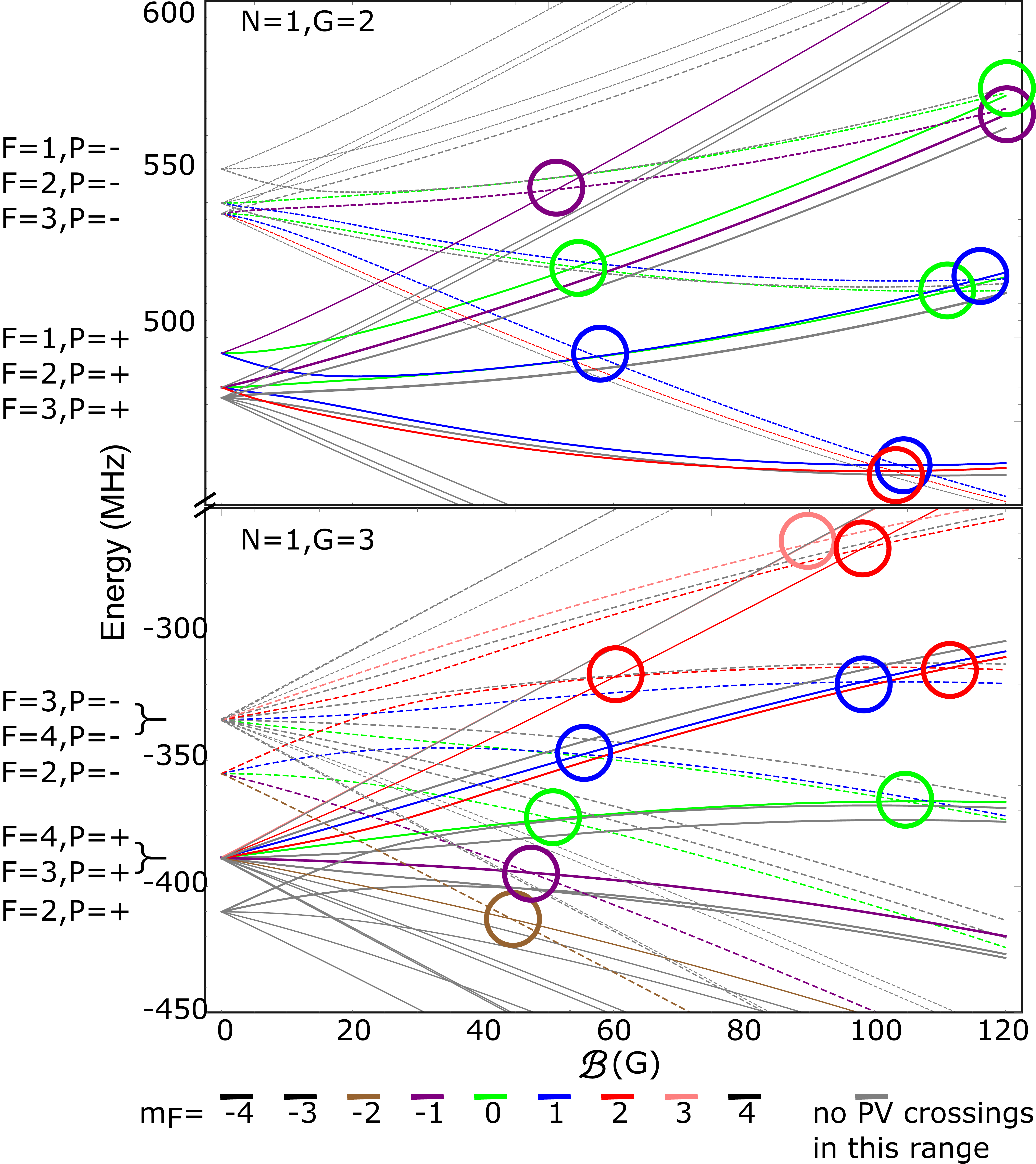}
   \caption{Examples of parity-violation-sensitive pair crossings in linear molecule.  Energy of MgNC $\ket{X^2\Sigma;(v_1 v_2^\ell v_3)=(01^10),N=1}$ spin-rotation and hyperfine Zeeman sublevels as a function of magnetic field $\mathcal{B}$.  Solid (dashed) lines have parity $P=+1$ ($P=-1$).  Crossings of parity-violation-sensitive pairs are circled. Gray lines correspond to state which do not have a parity-violation-sensitive crossing in this range of magnetic field. Other colors correspond to states with a particular value of total angular momentum project $m_F$ (see legend). Note that $\ket{N=1,G=3,F=3,4,P}$ are nearly degenerate when $\mathcal{B}=0$.} \label{fig:magic}
\end{figure}

\begin{figure*}[t]
    \centering
    \includegraphics[width=0.9\textwidth]{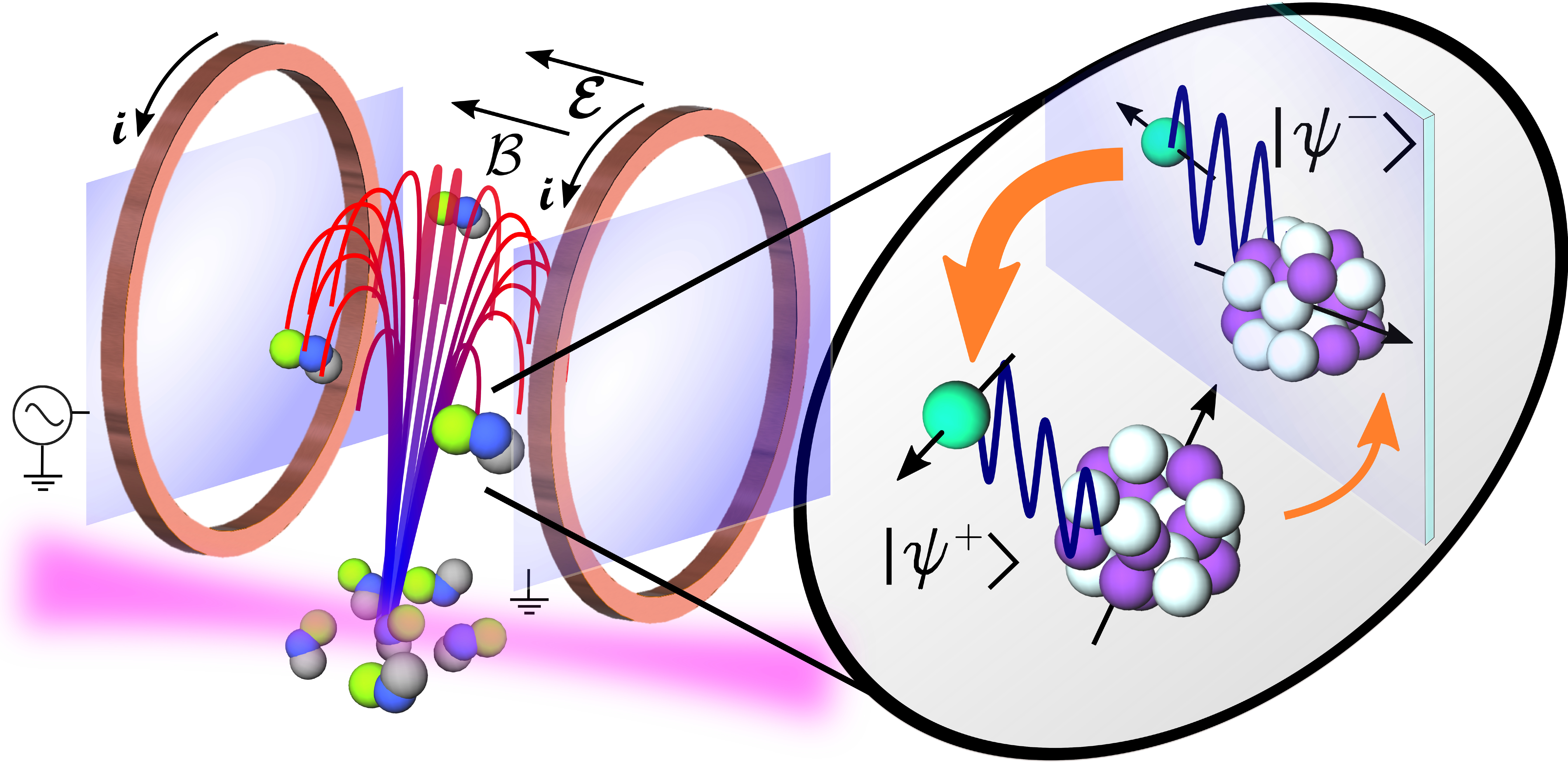}
    \caption{
    Potential nuclear spin-dependent parity-violation measurement scheme.  (Left) Laser-cooled triatomic molecules are prepared in the first bending mode to access the $\ell$-doublet structure, and are launched upward into an interaction region to form a molecule fountain. Oscillating electric field $\mathcal{E}$ drives electric dipole transitions between states of opposite parity. Magnetic field $\mathcal{B}$ tunes to degeneracy a particular pair of opposite-parity states $\ket{\psi^\pm}$ to enhance their interaction via the effective parity-violating Hamiltonian $H_{\text{NSD-PV}}^{\text{eff}}$. Population transfer from the initial state to the degenerate opposite-parity state is read out by laser spectroscopy after molecules fall back out of the interaction region. (Right) Stark interference: State transfer (orange) is parity dependent due to the combined NSD-PV interactions (wavy line) and electric dipole interaction interfering constructively or destructively  depending on the relative orientations of the electron spin, nuclear spin, and molecule axis.
    }
    \label{fig:fountain}
\end{figure*}

\subsection{Example: Parity-Violation-Sensitive Level Crossings in MgNC}
From Eq.\,(\ref{eq:asymmetry}), it is clear that the parity-violating asymmetry $\mathcal{A}$ is enhanced for small detuning $\Delta$.  In linear triatomic molecules, $\Delta$ can approach zero for parity-violation-sensitive pairs of levels via Zeeman tuning with modest magnetic field $\mathcal{B} \lesssim 100$\,G (1\,G\,=\,$10^{-4}$\,T).
Of the four molecular species considered here, MgNC has the most complete spectroscopy (Table~\ref{tab:MgNC}). The energy of  the $^{25}$MgNC measurement state (the $\ket{X^2\Sigma;(v_1 v_2^\ell v_3)=(01^10),N=1}$ state) is plotted as a function of magnetic field $\mathcal{B}$ in Fig.\,\ref{fig:magic}, with some simplifications and approximations discussed below.  Well over a dozen points are circled in Fig.\,\ref{fig:magic} identifying crossings of states which interact via Eq.\,(\ref{eq:effective hamiltonian}) in modest magnetic fields $\mathcal{B}\lesssim 100$\,G. Energies are calculated by diagonalizing the Hamiltonian Eq. (1) of Ref.\,\cite{norrgard2019nuclear}  using the measured spin-rotation and hyperfine constants presented in Table~\ref{tab:MgNC}.  Here, $B$ is the rotational constant, $\gamma$ is the spin-rotation constant, $b$ and $c$ are the Frosch and Foley hyperfine parameters, $q_{v=2}$ is the $\ell$-doubling constant, and $eq_0Q$ and $eq_2Q$ are electric quadrupole constants \cite{Brown2003,Hirota1985}.  Because $b \gg \gamma$, this state is best approximated by Hund's case ($b_{\beta S})$ in the field-free case. The states in Fig.\,\ref{fig:magic} are labeled according to the angular momentum coupling scheme $\boldsymbol{I}+\boldsymbol{S}=\boldsymbol{G},\boldsymbol{G}+\boldsymbol{N}=\boldsymbol{F}$.  Here $\boldsymbol{I}$ is the Mg nuclear spin ($I=5/2$), $\boldsymbol{S}$ is the total electron spin ($S=1/2$), $\boldsymbol{N}$ is the rotational plus orbital angular momentum, and $\boldsymbol{F}$ is the total angular momentum
\cite{Brown2003}.

\begin{table}\caption{Estimated spectroscopic parameters for the $\ket{X^2\Sigma, (v_1,v_2^\ell,v_3)=(01^10), N=1}$ state of $^{25}$MgNC.}
\begin{tabular}{llll}
\hline\hline
Parameter & Value (MHz) & State of Measurement                & Ref.              \\ \hline
$B$       & 6046.887    & $X^2\Sigma, (01^10)$ &\cite{Kagi1996}       \\
$\gamma$  & 15.3        & $X^2\Sigma, (01^10)$ &\cite{Kagi1996}       \\
$b$($^{25}$Mg)       & -298        & $X^2\Sigma, (00^00)$                                    &\cite{AndZiu94}   \\
$c$($^{25}$Mg)       & 14.72       & $X^2\Sigma, (00^00)$                                    &\cite{AndZiu94}   \\
$q_{v=2}$ & 27.3863     & $X^2\Sigma, (01^10)$  &\cite{Kagi1996}       \\
$eq_0Q$($^{25}$Mg)   & -19.5       & $X^2\Sigma, (00^00)$                                    &\cite{AndZiu94} \\
$eq_2Q$($^{25}$Mg)\quad\quad   & 40          & Estimate         &                       \\
\hline\hline
\end{tabular}\label{tab:MgNC}
\end{table}

In Fig.\,\ref{fig:magic}, the small hyperfine structure due to N and C is neglected \cite{WALKER1998}.  Inclusion of additional nuclear spins increases the number of states by a factor of the spin multiplicity, and in many case leads to additional level crossings.  For a $^2\Pi$ vibronic state, as is the case for the measurement state,  the T$^{2}_2(eqQ)$ component of the $^{25}$Mg electric quadrupole hyperfine constant $eq_2Q$ must be included \cite{Hirota1985}.  This parameter appears not to have been measured yet, but is typically of opposite sign and larger in magnitude than the parameter $eq_0Q$ \cite{Brown2003}. We use  $eq_2Q\,=\,40$\,MHz as a crude estimate.  Varying the value of $eq_2Q\,=\,0$\,MHz to 100\,MHz in our calculation is found to not change general features of of Fig.\,\ref{fig:magic}, only the values of $\mathcal{B}$ at which levels cross.  The hyperfine parameters taken from Ref.\,\cite{WALKER1998} were measured in the $(v_1 v_2^\ell v_3)=(00^00)$ vibrational state, but due to the ionic nature of the Mg-N bond should not change substantially when exciting to the first bending mode.  Finally, the parameters listed in Table~\ref{tab:MgNC} result in the levels $\ket{N=1,G=3,F=3,4,P}$ being nearly degenerate ($\sim$\,1\,MHz spacing) in the field-free case; this is insubstantial to the NSD-PV measurement but is noted for clarity of Fig.\,\ref{fig:magic}.

\subsection{Free Flight Method}

In the free flight method, laser-cooled molecules may be launched upward into an interaction region (e.g.\ via Stark acceleration \cite{Cheng2016} or white light pushing laser beams \cite{Barry2012}) to form a molecule fountain as illustrated in Fig.\,\ref{fig:fountain}, or more simply, released  to drop through an interaction region.  Once in free flight, molecules are efficiently prepared in a particular Zeeman sublevel of the $\ket{X^2\Sigma;(v_1 v_2^\ell v_3)=(01^10),N=1}$ manifold via Stimulated Raman adiabatic passage \cite{Gaubatz1990} in order to access the $\ell$-doublet structure.   The interaction region contains electric field plates to provide the oscillating electric field $\mathcal{E}$, and electromagnet coils to Zeeman tune to degeneracy a particular pair of opposite-parity states. The population transfer signal $S$ is read out by laser spectroscopy as molecules fall out of the interaction region. 

The primary advantage of the free flight method compared to the optical trapping scheme of Ref.\,\cite{norrgard2019nuclear} is the potential to increase interaction time $\tau$. This is at the expense of increasing the volume $V$ explored by the molecules during the interaction time.  The optimal interaction time (and therefore the measurement sensitivity) will be limited by the magnitude of uncompensated electric and magnetic fields and field gradients over the interaction volume $V$.  For a molecule ensemble with typical velocity $v_t$ and initial spatial extent $r$, when $\tau \ll r/v_{\text{t}}$, $V$ is proportional to $\tau^2$.  In the case $\tau \gg r/v_{\text{t}}$ is proportional to $\tau^4$.  In contrast, for molecules in an optical trap the field must only be homogeneous over a small interaction volume of typically $\lesssim 1 $\,mm$^3$ regardless of interaction time. 
A typical molecule magneto-optical trap has radius $r\sim\,2$\,mm \cite{Norrgard2016,Truppe2017}.  The optical cycling transition of MgNC has a wavelength of $\lambda$\,=\,383\,nm \cite{Kagi1996,WALKER1998}, corresponding to a recoil velocity $v_r\,=\,h /m\lambda\,=\,2.0$\,cm/s and recoil temperature $T_r\,=\,m v_r^2/2 k_{\text{B}}$\,=\,2.5\,$\mu$K ($m$ is the molecule mass and $k_{\text{B}}$ is the Boltzmann constant).  With an  interaction time $\tau=10$\,ms and assuming modest sub-Doppler cooling to $T\,=\,10T_r$, the molecule cloud would expand $\sim 0.6$\,mm in each dimension while dropping $0.5$\,mm (or rising  0.25\,mm and returning in the case of a fountain).  In this case the interaction volume could be limited to $V\,\lesssim\,20$\,mm$^3$.

Assuming $\tau\,=\,10$\,ms for a free flight measurement, we may estimate the sensitivity to NSD-PV effects. Considering typical for diatomic molecule magneto-optical trap populations and assuming good launch efficiency into the interaction region, we expect  $N\, \simeq \,10^4$ molecules per measurement.  Compared to the case of loading an optical dipole trap, we expect more molecules in a free flight measurement as it avoids additional losses from inefficient trap transfers.  The repetition rate will likely be somewhat reduced due to the longer measurement sequence to $\mathcal{R}\,\simeq \,1$\,s$^{-1}$.  The rate of molecules measured $\mathcal{R}N$ is therefore likely to be comparable between the two methods. For $\tau\,=\,10$\,ms, we expect $\delta W\,=\,2\pi\times0.16$\,Hz$/\sqrt{\text{Hz}}$ for a free-flight NSD-PV measurement, or roughly a factor of 6 more sensitive than the optical dipole trapping scheme.  

Achieving interaction time-limited sensitivity at $\tau\,=\,10$\,ms  would require a corresponding 6-fold improvement in the magnetic field uniformity over the entire interaction volume; in this example, $\delta \mathcal{B} \lesssim 1/\tau \mu_{\text{B}} \simeq 70\,\mu$\,G (with $\mu_{\text{B}}$ the Bohr magneton).  The low magnetic fields required for this measurement may easily be reversed to aid in detecting field gradients.  $\mathcal{B}$-field mimicing $\mathcal{E}\times \textbf{v}$ effects from nonreversing electric fields $\mathcal{E}_{\text{nr}}$ are unlikely to be important; for example, $\mathcal{E}_{\text{nr}}=\,1\,$mV/cm corresponds to a magnetic field uncertainty of $\simeq 300$\,nG after gravitational acceleration  for $\tau\,=\,10$\,ms.  

 Finally, we may comment on the expected magnitude of known systematic effects which mimic the NSD-PV asymmetry signal.  In the recent measurements in a BaF beam,  the combination of residual $\mathcal{B}$-field gradient with $\mathcal{E}_{\text{nr}}$ was shown to mimic the NSD-PV asymmetry at the 10\,mHz level\cite{AltAmmCah18-1,AltAmmCah18-2}.  As the required $\mathcal{B}$ is reduced by a factor of roughly 100--1000 compared to diatomic molecules, if we assume the gradients are proportional to the magnitude of the field applied, this systematic uncertainty would appear at the 10\,$\mu$Hz--100\,$\mu$Hz level for polyatomic molecules. Using as a typical value of $\Tilde{C}\equiv\langle(\bm{\hat{n}}\times \bm{S}_{\text{eff}} )\cdot \bm{I}/I\rangle \simeq 0.1$ \cite{demille2008using,norrgard2019nuclear}, this systematic error is below the total theoretical uncertainty of $W = \kappa W_{\text{PV}}\Tilde{C}$ for $^{25}$Mg, but will need to be reduced by roughly an order of magnitude when measuring NSD-PV in the other nuclei.

\section{Conclusions}
 
We have calculated for multiple nuclei and molecular species the parameters necessary to interpret NSD-PV measurements in light triatomic molecules: nuclear parameter $\kappa$ with $\sim 30\%$ uncertainty and molecular parameter $W_{\text{PV}}$ with 5\%-10$\%$ uncertainty. The enhanced sensitivity to NSD-PV effects in polyatomic molecules enables measurements in these light species, where the confluence of our molecular and nuclear calculations allows us to predict the expected Standard Model NSD-PV effect with unprecedented high accuracy. This is a marked improvement in theoretical uncertainty compared to heavy atoms and molecules.

By using laser-cooled polyatomic molecules in a fountain-like configuration, we have estimated that the matrix elements of Eq.\,(\ref{eq:effective hamiltonian}) could be measured with an experimental sensitivity of roughly $2\pi\times 0.16$ $\text{Hz}/\sqrt{\text{Hz}}$.  
At this sensitivity, the $W_{\text{PV}}$ values and Standard Model values for $\kappa$ calculated here suggest that a  10\,\%  statistical uncertainty measurement could be achieved with an averaging time of about 1 hour for $^{25}\text{Mg}$, or about one week for $^{9}$Be, $^{13}$C, and $^{14,15}$N. Should it turn out that the NSD-PV effect is significantly larger than expected, it would potentially be a signature of new  physics,  such as a $Z^\prime$ boson.

\section*{Acknowledgement}
The authors would like to thank the Center for Information Technology (CIT) of the University of Groningen for providing access to the Peregrine high-performance computing cluster and for their technical support. We thank I. Stetcu for useful discussions and B. A. Brown for useful discussions and benchmarking of the PNC NN matrix elements. P.N. acknowledges support from the NSERC Grant No. SAPIN-2016-00033. TRIUMF receives federal funding via a contribution agreement with the National Research Council of Canada.  Computing support came from an INCITE Award on the Summit supercomputer of the Oak Ridge Leadership Computing Facility (OLCF) at ORNL, and from Westgrid and Compute Canada.
M.I. acknowledges the support of the Slovak Research and Development Agency and the Scientific Grant Agency, Grants No. APVV-15-0105 and No. VEGA 1/0562/20, respectively.  E.B.N. was supported by NIST and the National Research Council Postdoctoral Research
Associateship Program, and thanks Stephen Eckel for careful reading of the manuscript. This research used the resources of a High Performance Computing Center of the Matej Bel University in Banska Bystrica using the HPC infrastructure acquired in Projects No. ITMS 26230120002 and No. 26210120002 (Slovak infrastructure for high performance computing) supported by the Research and Development Operational Programme funded by the ERDF. V.F. acknowledges partial support from the Australian Research Council, Gutenberg Fellowship and New Zealand Institute for Advanced Studies. 

\appendix
\section*{Appendix: Signs and Notation}

Here we clarify some confusion in the sign of the $\bm{V}_e\bm{A}_N$  parameter (here called $\kappa_{\text{ax}}$) in recent literature, using $^{133}$Cs as an example.
All sources presented here predict a positive contribution for the $\bm{V}_e\bm{A}_N$  term to $\kappa$(Cs), except for \cite{demille2008using} and \cite{norrgard2019nuclear}.  
The discrepancy appears to be due to is an incorrect substitution in  Refs. \cite{demille2008using} and \cite{norrgard2019nuclear} of $C_2\rightarrow C_{2\nu}$, instead of the  $C_2\equiv -C_{2\nu}$ used elsewhere.  The subtlety of this point is compounded by the inconsistency in the literature on the symbol for this ($C_2$, $K_2$, $\kappa_2$ have all been used) and a proliferation of minus signs.  
Various examples of notations for $\kappa$, definition of $C_2$, and the $\bm{V}_e\bm{A}_N$ contribution to $\kappa$ and its sign are listed
 below.  Note that for Cs,  we have a valence proton, $I=7/2$, $\ell_\nu$=4, $K=4$.
 
This work:
  \begin{equation}
  \begin{aligned}
    &\kappa = \kappa_\text{A} + \kappa_\text{ax} +\kappa_\text{hfs},\\
  &\kappa_{\text{ax}} = C_{2} \frac{1/2-K}{I+1}, \\
  &C_2 = -C_{2P} =-0.05, \\
  &\kappa_\text{ax}(\text{Cs}) > 0.
\end{aligned}
\end{equation}

Flambaum \& Khriplovich \cite{FlaKhr80-1}:
 \begin{equation}
  \begin{aligned}
  &\kappa = \kappa_\text{a} + \kappa_\text{ax} +\kappa_\text{hfs}\\
  &\kappa_{\text{ax}} = - \frac{K-1/2}{I+1}K_2, \\
  &K_2= -C_{2P} =-0.05, \\
  &\kappa_\text{ax}(\text{Cs}) > 0.
  \end{aligned}
\end{equation}

Flambaum \& Murray \cite{FlaMur97}:
 \begin{equation}
  \begin{aligned}
  &\kappa = \kappa_\text{a} - \kappa_\text{ax} +\kappa_\text{hfs}\\
  &\kappa_{\text{ax}} = - \frac{K-1/2}{K}\kappa_2, \\
  &\kappa_2=-C_{2P} =-0.05, \\
  &-\kappa_\text{ax}(\text{Cs}) > 0.
\end{aligned}
\end{equation}

Haxton \& Wieman \cite{haxton2001atomic}
 \begin{equation}
  \begin{aligned}
  &\kappa = \kappa_{\text{anapole}} + \kappa_{Z^0} +\kappa_{Q_W}\\
  &\kappa_{Z^0} = - 1.26(1-4\sin^2\theta_W)\frac{(-1)^{I-\ell-m_t}}{2\ell+1}, \\
  &m_t =+1/2\ (-1/2)\ \text{for proton (neutron)}\\
  &\kappa_{Z^0}(\text{Cs}) = +0.0140 > 0.
  \end{aligned}
\end{equation}

Gringes \& Flambaum \cite{ginges2004violations}:
\begin{equation}
  \begin{aligned}
  &\hat{h}=\hat{h}_{NC}+\hat{h}_{NC}^I+\hat{h}^I_Q\\
  &\hat{h}^I_{NC}= - \kappa_2 \frac{K-1/2}{I+1}, \\
  &\kappa_2= -C_{2},\\
  &C_2 = C_{2P} = 0.05, \\
  &\hat{h}^I_{NC}(\text{Cs}) > 0.
  \end{aligned}
\end{equation}

Gomez \textit{et al.} \cite{Gomez2007}:
 \begin{equation}
  \begin{aligned}
  &\kappa = \kappa_a -\frac{K-1/2}{K} \kappa_{2} +\frac{I+1}{K}\kappa_{Q_W}\\
  &\kappa_{2,p} = - C_{2P}= -0.05, \\
  &-\frac{K-1/2}{K}\kappa_{2,p}> 0.
\end{aligned}
\end{equation}

DeMille \textit{et al.} \cite{demille2008using} and Norrgard \textit{et al.} \cite{norrgard2019nuclear}:
 \begin{equation}
  \begin{aligned}
  &\kappa^\prime = \kappa_a^\prime + \kappa_{2}^\prime +\kappa_{Q}^\prime\\
  &\kappa_{2}^\prime = C_{2\nu} \frac{1/2-K}{I+1}, \\
  &C_{2P} = +0.05, \\
  &\kappa_{2}^\prime(\text{Cs}) < 0.
  \end{aligned}
\end{equation}
While neither of Refs. \cite{demille2008using} and \cite{norrgard2019nuclear} list $\kappa_{2}^\prime(\text{Cs})$ explicitly, we can check the sign using Table I of Ref.\,\cite{demille2008using} for $^{139}$La, as it matches the case for $^{133}$Cs (valence proton, $I=7/2$, $\ell$=4, $K=4$); they give $\kappa_2^\prime= -0.039$.

\clearpage

\bibliography{triatomic}
\end{document}